\documentclass[twocolumn,showpacs,preprintnumbers,
amsmath,amssymb,aps,prd,nofootinbib,superscriptaddress,
eqsecnum]{revtex4}
\usepackage{graphicx}

\makeatletter
\renewcommand{\p@subsection}{}
\makeatother

\renewcommand{\thesection}{\arabic{section}}

\renewcommand{\theequation}{\arabic{section}.\arabic{equation}}

\newcommand{\chushi}[1]{ }

\begin{document}

\title{Enhancement of quark number susceptibility with an 
alternative pattern of chiral symmetry breaking in dense matter}

\author{Masayasu Harada}
\affiliation{%
Department of Physics, Nagoya University,
Nagoya, 464-8602, Japan
}
\author{Chihiro Sasaki}
\affiliation{%
Physik-Department,
Technische Universit\"{a}t M\"{u}nchen,
D-85747 Garching, Germany
}
\author{Shinpei Takemoto}
\affiliation{%
Department of Physics, Nagoya University,
Nagoya, 464-8602, Japan
}

\date{\today}

\begin{abstract}
We explore a possible phase where chiral
$SU(2)_L \times SU(2)_R$ symmetry is spontaneously broken 
while its center $Z_2$ symmetry remains unbroken and 
its consequence on thermal quantities.
In this phase, chiral symmetry breaking is driven 
by a quartic quark condensate although a bilinear 
quark condensate vanishes.
A Ginzburg-Landau free energy leads to
a new tricritical point (TCP) between the $Z_2$ 
broken and unbroken phases.
Furthermore, a critical point can appear even in the
chiral limit where explicit breaking is turned off,
instead of a TCP at which restoration of chiral and 
its center symmetries takes place simultaneously.
The net quark number density exhibits an abrupt change 
near the restoration of the center symmetry rather than that 
of the chiral symmetry.
Hadron masses in possible phases are also studied in a linear 
sigma model. 
We show that, in the $Z_2$ symmetric phase,
the $\bar{q}q$-type scalar meson with zero isospin $I=0$
splits from the $\bar{q}q$-type pseudoscalar meson with
$I=1$.
\end{abstract}

\pacs{11.30.Rd, 11.30.Ly, 25.75.Nq, 21.65.Qr}

\maketitle

\section{Introduction}
\label{sec:intro}

Properties of hot and/or dense QCD matter has been extensively 
studied within chiral approaches~\cite{review}.
Our knowledge on the phase structure is however still limited
and the description of the matter around the phase transitions 
does not reach a consensus, where a typical
size of the critical temperature and chemical potential is
considered to be of order $\Lambda_{\rm QCD}$.
The phases of QCD are characterized by symmetries and their
breaking pattern: 
QCD at asymptotically high density leads to the color-flavor-locked
phase as the true ground state under the symmetry breaking pattern,
$SU(3)_c \times SU(3)_L \times SU(3)_R$ down to the diagonal
subgroup $SU(3)_{c + L + R}$~\cite{cfl}. The residual discrete
symmetries characterize the spectra of excitations.

At zero temperature and density,
an alternative pattern of spontaneous chiral symmetry breaking 
was suggested in the context of QCD~\cite{stern,SDE,SDE2}.
This pattern keeps the center of chiral group unbroken, i.e.
$SU(N_f)_L \times SU(N_f)_R \to SU(N_f)_V \times (Z_{N_f})_A$,
where a discrete symmetry $(Z_{N_f})_A$ is the maximal axial
subgroup of $SU(N_f)_L \times SU(N_f)_R$. The $Z_{N_f}$ symmetry
protects a theory from condensate of quark bilinears 
$\langle \bar{q}q \rangle$. Spontaneous symmetry breaking
is driven by quartic condensates which are invariant under
both $SU(N_f)_V$ and $Z_{N_f}$ transformation.
Although meson phenomenology with this breaking pattern
seems to explain the reality reasonably~\cite{stern},
this possibility is strictly ruled out in QCD both at zero
and finite temperatures but at zero density since a different
way of coupling of Nambu-Goldstone bosons to pseudo-scalar
density violates QCD inequalities for density-density
correlators~\cite{shifman}.
However, this does not exclude the unorthodox pattern
in the presence of dense baryonic matter. There are several 
attempts which dynamically generate a similar breaking pattern 
in an O(2) scalar model~\cite{fukushima} and in ${\cal N}=1$
Super Yang-Mills theory~\cite{SYM}.

Within the Skyrme model on crystal,
a new intermediate phase where a skyrmion turns into two half 
skyrmions was numerically found~\cite{skyrmion}.
This phase is characterized by a vanishing quark condensate 
$\langle\bar{q}q \rangle$ and a non-vanishing pion decay constant.
Recently, another novel view of dense matter, Quarkyonic Phase, 
has been proposed based on the argument using the large $N_c$ 
counting where $N_c$ denotes the number of colors~\cite{quarkyonic}:
In the large $N_c$ limit there are three phases which are
rigorously distinguished using the Polyakov loop expectation
value $\langle \Phi \rangle$ and the baryon number density
$\langle N_B \rangle$. The quarkyonic phase is characterized
by $\langle \Phi \rangle = 0$ indicating the system confined
and non-vanishing $\langle N_B \rangle$ above $\mu_B = M_B$
with a baryon mass $M_B$. The separation of the quarkyonic
from hadronic phase is not clear any more in a system with 
finite $N_c$. Nevertheless, an abrupt change in the baryon number
density would be interpreted as the quarkyonic transition which 
separates meson dominant from baryon dominant regions. This might 
appear near the boundary for chemical equilibrium
at which one would expect a rapid change in the number of degrees 
of freedom~\cite{quarkyonic,larry}.

A steep increase in the baryon number density and the corresponding 
maximum in its susceptibility $\chi_B$ are driven by a phase transition 
from chirally broken to restored phase in most model-approaches.
Interplay between (de)confinement and chiral symmetry breaking
has been studied within a Nambu--Jona-Lassinio model with Polyakov 
loops~\cite{pnjl} which describes how the deconfinement and 
chiral phase boundaries are changed from $N_c = \infty$ down to 
$N_c=3$~\cite{mrs:pnjl}. The model study shows that the chiral
phase transition at $T=0$ appears just above mass threshold 
$\mu_B = M_B$ and thus a large $\chi_B$ is associated with
the chiral phase transition.
However, a constituent-quark picture does not directly describe
the thermodynamics of hadronic matter
and there are no {\it a priori} reasons that the quarkyonic
transition should be accompanied by chiral phase transition.
Besides, it seems unlikely that the chiral symmetry is (even
partially) restored slightly above the freeze-out curve where
the baryon density is not high enough to drive a phase transition.
{}From this perspective, further investigations of dense baryonic
matter and a possible appearance of the quarkyonic phase
in QCD with $N_c = 3$ require a modeling in terms of dynamical
hadronic-degrees of freedom in a systematic way.

In this paper we will address this issue under the alternative
pattern of chiral symmetry breaking in dense hadronic matter.
We will show a possible intermediate phase between
chiral symmetry broken and its restored phases
with analyses using a general Ginzburg-Landau free energy.
This leads to multiple critical points and one of them
is associated with restoration of the center symmetry
rather than that of chiral symmetry.
In the new phase with unbroken center symmetry
the net baryon number susceptibility exhibits a strong enhancement
although the chiral symmetry remains spontaneously broken.
This is reminiscent of the quarkyonic transition and our framework
provides a theoretical description of the quarkyonic phase
on the bases of a chiral Lagrangian with two distinct order parameters. 
An analysis using a linear sigma model for hadron mass spectra 
is also made.

\section{A model for 2-quark and 4-quark states}
\label{sec:model}

We construct a chiral Lagrangian for 2- and 4-quark states
under the following pattern of symmetry breaking,
\begin{eqnarray}
SU(N_f)_L \times SU(N_f)_R 
&\to& SU(N_f)_V \times (Z_{N_f})_A
\nonumber\\
&\to& SU(N_f)_V\,.
\end{eqnarray}
In this paper we will restrict ourselves to a two-flavor case.

\subsection{Lagrangian}
\label{ssec:lag}

We introduce a 2-quark state $M$ in the fundamental
and a 4-quark state $\Sigma$ in the adjoint representation 
as~\footnote{
 We consider $\Sigma$ as any linear combination of
 $\bar{q}q$-$\bar{q}q$ and $\bar{q}\bar{q}$-$qq$ type fields
 allowed by symmetries.
}
\begin{eqnarray}
M_{ij} &\sim& \bar{q}_{R,j} q_{L,i}\,,
\nonumber\\
\Sigma_{ab} &\sim& 
    \bar{q}_{L} \tau_a \gamma_\mu q_L
    \bar{q}_{R} \tau_b \gamma^\mu q_R\,,
\end{eqnarray}
where the flavor indices run $(i,j) = 1,2$ and
$(a,b,c) = 1,2,3$ and Pauli matrices $\tau^a = 2 T^a$ with
$\mbox{tr}[T^a T^b] = \delta^{ab}/2$.
The $M$ and $\Sigma$ are expressed as
\begin{eqnarray}
M_{ij} &=& \frac{1}{\sqrt{2}}\left( \sigma\delta_{ij} 
{}+ i\phi^a\tau^a_{ij} \right)\,,
\nonumber\\
\Sigma_{ab} &=& \frac{1}{\sqrt{3}}\chi\delta_{ab}
{}+ \frac{1}{\sqrt{2}}\epsilon_{abc}\psi_c\,,
\end{eqnarray}
where $\sigma$ and $\chi$ represent scalar fields and 
$\phi$ and $\psi$ pseudoscalar fields, and $\epsilon_{ijk}$
is the total anti-symmetric tensor with $\epsilon_{123} = 1$.
In general the field $\Sigma$ contains an isospin 2 state.
One can take appropriate parameters in a Lagrangian in such
a way that this exotic particle is very heavy.
Thus, we will consider only isospin 0 ($\chi$) and 1 ($\psi$) 
states in this paper.
The fields transform under $SU(2)_L\times SU(2)_R$ 
as chiral non-singlet,
\begin{eqnarray}
M \to g_L^{(2)} M\, g_R^{(2)\dagger}\,,
\quad
\Sigma \to g_L^{(3)} \Sigma\, g_R^{(3)\dagger}\,.
\end{eqnarray}
This transformation property implies that the field
$M$ changes its sign under the center $Z_2$ of SU(2)$_L$ 
(or SU(2)$_R$), while $\Sigma$ is invariant:
\begin{eqnarray}
M \to -M\,,
\quad
\Sigma \to \Sigma\,.
\end{eqnarray}
Up to the fourth order in fields one obtains a potential,
\begin{eqnarray}
&&
V(M,\Sigma) 
= -\frac{m^2}{2}\mbox{Tr}\left[ M M^\dagger\right]
{}+ \frac{\lambda^2}{4}\left(\mbox{Tr}\left[M M^\dagger\right]
\right)^2
\nonumber\\
&&
{}- \frac{\bar{m}^2}{2}\Sigma_{ab}\Sigma^T_{ba}
{}+ \frac{\bar{\lambda}_1^2}{4}\Sigma_{ab}\Sigma^T_{bc}
\Sigma_{cd}\Sigma^T_{da}
{}+ \frac{\bar{\lambda}_2^2}{4}
\left( \Sigma_{ab}\Sigma^T_{ba} \right)^2
\nonumber\\
&&
{}+ 2g_1\Sigma_{ab}\mbox{Tr}\left[ T_a M T_b M^\dagger \right]
{}+ g_2\Sigma_{ab}\Sigma^T_{ba}\mbox{Tr}\left[ M M^\dagger\right]
\nonumber\\
&&
{}+ g_3 \mbox{Det}\Sigma 
{}+ g_4 (\mbox{Det} M + \mbox{h.c.})\,.
\label{potential}
\end{eqnarray}
The last term violates the $U(1)_A$ symmetry.
The coefficients of the quartic terms are positive for this 
potential to be bounded. Other parameters $g_i$ 
can be both positive and negative and will determine the topology
of the phase structure.
An explicit chiral symmetry breaking can be introduced 
through, e.g.,
\begin{equation}
V_{\rm SB}(M,\Sigma)
= -h\sigma - \alpha h^2\chi\,,
\end{equation}
with constants $h$ and $\alpha$.
Note that a similar Lagrangian was considered for a system
with 2- and 4-quark states under the symmetry breaking
pattern without unbroken center symmetry in~\cite{rischke} where 
their 4-quark states are chiral singlet and the potential
does not include quartic terms in fields.

\subsection{Ginzburg-Landau effective potential}
\label{ssec:gl}

We first study possible phases derived from the effective
potential~(\ref{potential}) taking
\begin{equation}
M_{ij} = \frac{1}{\sqrt{2}}\sigma\delta_{ij}\,,
\quad
\Sigma_{ab} = \frac{1}{\sqrt{3}}\chi\delta_{ab}\,.
\end{equation}
One can reduce Eq.~(\ref{potential}) as well as an explicit 
breaking term to
\begin{eqnarray}
V(\sigma,\chi)
&=& 
A\sigma^2 + B\chi^2 + \sigma^4 + \chi^4 - h\sigma
\nonumber\\
&&
{}+ C\sigma^2\chi + D\chi^3 + F\sigma^2\chi^2\,.
\label{gl}
\end{eqnarray}
We will take $C=-1$ without loss of generality 
in the following calculations.

\begin{figure}
\begin{center}
\includegraphics[width=7cm]{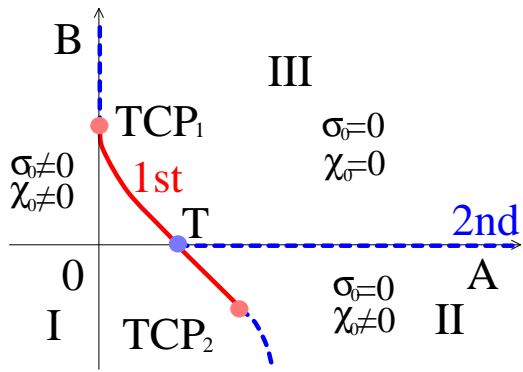}
\caption{
Phase diagram with $D=F=0$ and $h=0$.
The solid and dashed lines indicate first and second order 
phase boundaries, respectively.
One tricritical point, TCP$_1$, is located at
$(A,B)=(0,1/4)$ and another, TCP$_2$, at $(A,B)=(1/4,-1/8)$.
The triple point represented by $T$ is at $(A,B)=(1/8,0)$.
}
\label{d0}
\end{center}
\end{figure}

We start with the potential for $D=F=0$ and $h=0$,
\begin{equation}
V = A\sigma^2 + B\chi^2 + \sigma^4 + \chi^4 - \sigma^2\chi\,.
\end{equation}
Phases from this potential can be classified by the
coefficients $A$ and $B$. The expression of the phase boundaries
is summarized in Appendix~\ref{app:phase}.
Here we discuss the obtained phase structure shown in 
Fig.~\ref{d0}.
There are three distinct phases characterized by two order
parameters: Phase I represents the system where both chiral 
symmetry and its center are spontaneously broken due to 
non-vanishing expectation values $\chi_0$ and $\sigma_0$. 
The center symmetry
is restored when $\sigma_0$ becomes zero. However, chiral
symmetry remains broken as long as one has non-vanishing $\chi_0$,
indicated by 
phase II. The chiral symmetry restoration takes
place under $\chi_0 \to 0$ which corresponds to phase III.
The phases II and III are separated by a second-order line,
while the broken phase I from II or from III is by both first-
and second-order lines. 
Accordingly, 
there exist two tricritical points
(TCPs) and one triple point. One of these TCP, TCP$_2$ 
in Fig.~\ref{d0}, is associated with the center $Z_2$ symmetry
restoration rather than the chiral transition.

Two phase transitions are characterized by susceptibilities
of the corresponding order parameters. We introduce a 2-by-2 matrix 
composed of the second derivatives of $V$ as
\begin{equation}
\hat{C} = \left( 
\begin{matrix}
C_{\sigma\sigma} & C_{\sigma\chi}
\\
C_{\chi\sigma} & C_{\chi\chi}
\end{matrix}
\right)\,,
\end{equation}
with
\begin{eqnarray}
&&
C_{\sigma\sigma} = \frac{\partial^2 V}{\partial\sigma^2}\,,
\quad
C_{\chi\chi} = \frac{\partial^2 V}{\partial\chi^2}\,,
\nonumber\\
&&
C_{\sigma\chi} = C_{\chi\sigma}
= \frac{\partial^2 V}{\partial\sigma\partial\chi}\,,
\end{eqnarray}
under the solutions of the gap equations, $\sigma_0$
and $\chi_0$.
A set of susceptibilities is defined by the inverse of
$\hat{C}$~\cite{sfr:pnjl};
\begin{equation}
\hat{\chi} = \frac{1}{\mbox{det}\hat{C}}
\left( 
\begin{matrix}
C_{\chi\chi} & -C_{\sigma\chi}
\\
-C_{\chi\sigma} & C_{\sigma\sigma}
\end{matrix}
\right)\,.
\end{equation}
We identify the susceptibilities associated with
2-quark and 4-quark states as
\begin{equation}
\chi_{\rm 2Q} = \hat{\chi}_{11}\,,
\quad
\chi_{\rm 4Q} = \hat{\chi}_{22}\,.
\end{equation}
The $\chi_{\rm 2Q}$ is responsible to the $Z_2$ symmetry 
and the $\chi_{\rm 4Q}$ to the chiral symmetry restoration.

We consider $\chi_{\rm 2Q}$ and $\chi_{\rm 4Q}$ around 
the TCP$_1$ in Fig.~\ref{d0} where the potential has
zero curvature and thus $\mbox{det}\hat{C} = 0$.
When approaching the TCP$_1$ from broken phase I by
tuning $A$ and $B$ as $A \rightarrow A_{\rm critical} = 0$
and $B=1/4$,
these susceptibilities diverge as
\begin{equation}
\chi_{\rm 2Q} \sim t^{-1}\,,
\quad
\chi_{\rm 4Q} \sim t^{-2/3}\,,
\end{equation}
where $A_{\rm critical} - A \sim t$ with the reduced temperature 
or chemical potential,
e.g. $t=\lvert \mu-\mu_c \rvert/\mu_c$.
The gap equations determine
the scaling of 2-quark and 4-quark condensates as 
\begin{equation}
\sigma_0^2 \sim t^{1/3}\,, \quad
\chi_0 \sim t^{1/3}
\,.
\end{equation}
Consequently,
the quark number susceptibility 
$\chi_q = -\partial^2V/\partial\mu^2$ exhibits a singularity
as
\begin{equation}
\chi_q \sim \sigma_0^2\cdot \chi_{\rm 2Q} \sim t^{-2/3}\,.
\label{chi q TCP1}
\end{equation}
This critical exponent is same as the one
in the 3-d Ising model.
The coincidence can be understood due to the same $Z_2$ 
symmetries~\footnote{
 The $Z_2$ symmetry in the 3-d Ising system is not the center 
 of two-flavor chiral group, but emerges in the direction of 
 a linear combination of
 quark number and scalar densities~\cite{fujii}.
}.

The critical behavior near the TCP$_2$ involves more:
When the $A$ is approached as $1/4 - t$ with $B=-1/8$ fixed,
$\chi_{\rm 2Q}$ and $\chi_{\rm 4Q}$ diverge as
\begin{equation}
\chi_{\rm 2Q} \sim t^{-1}\,,
\quad
\chi_{\rm 4Q} \sim t^{-1/2}\,,
\end{equation}
and
only $\sigma_0$ vanishes as
$\sigma_0^2 \sim t^{1/2}$.
As a result, the quark number susceptibility $\chi_q$ 
diverges as
\begin{equation}
\chi_q \sim t^{-1/2}\,.
\end{equation}
Note that the critical exponent $1/2$ is different from the
one near the TCP$_1$,
which may reflect different symmetries possessed by the system
at TCP$_2$, $SU(2)_V$ and the center $Z_2$, from that at TCP$_1$,
$SU(2)_L \times SU(2)_R$ 
including its center $(Z_2)_L \times (Z_2)_R$.
Those exponents at TCP$_{1,2}$ are changed when $D \neq 0$
 (see below).

When the second-order phase transition separating
phase I from II or from III is approached from the broken 
phase with a fixed $B$, we have
\begin{equation}
\chi_{\rm 2Q} \sim t^{-1}\,,
\quad
\chi_{\rm 4Q} \sim \frac{1}{B}\,,
\end{equation}
where $B$ is a finite number, which thus gives 
no singularities in $\chi_{\rm 4Q}$.
The 2-quark condensate scales as $\sigma_0^2 \sim t^1$
and the quark number susceptibility $\chi_q$ is finite along
the second-order phase transition line:
\begin{equation}
\chi_q \sim \sigma_0^2\cdot \chi_{\rm 2Q} \sim t^0\,.
\end{equation}
Nevertheless, $\chi_q$ is enhanced toward the phase
transition induced by $\chi_{\rm 2Q}$ and becomes small
above the phase transition.
Such abrupt changes in $\chi_q$ indicate the phase transition,
especially for a negative $B$ which is driven by the center 
symmetry restoration rather than the chiral phase transition.

Near the second-order chiral transition between phase II
and III,
one obtains from $B \sim t$
\begin{equation}
\chi_0^2 \sim t^1\,,
\quad
\chi_{\rm 4Q} \sim t^{-1}\,.
\end{equation}
Since the chiral symmetry including the center symmetry 
prohibits the Yukawa-type coupling
of $\chi$ to a fermion and an anti-fermion in the fundamental
representation, 
the coupling of $\chi$ to the baryon number current would
be highly suppressed.
Therefore, $\chi_q$ shows less sensitivity 
around the chiral transition.~\footnote{
 As we will show below, the phase transition from phase II
 to phase III is of first order in a more general parameter choice.
 Thus, $\chi_q$ exhibits a jump at the chiral phase transition point.
}

Once small $h$ is turned on, chiral symmetry and its
center are explicitly broken. Second-order phase boundaries are
replaced with cross over and the two TCPs with two critical
points. The singularity in $\chi_q$ is now governed by
the $Z_2$ universality class of 3-d Ising systems. 
Thus, the scaling of $\chi_q$ at the critical points (CPs) 
will be given by
\begin{equation}
\chi_q \sim t^{-2/3}\,.
\end{equation}

\begin{figure*}
\begin{center}
(i) $-1 < D < 0$
\hspace*{6cm}
(ii) $D \leq -1$
\\
\includegraphics[width=7cm]{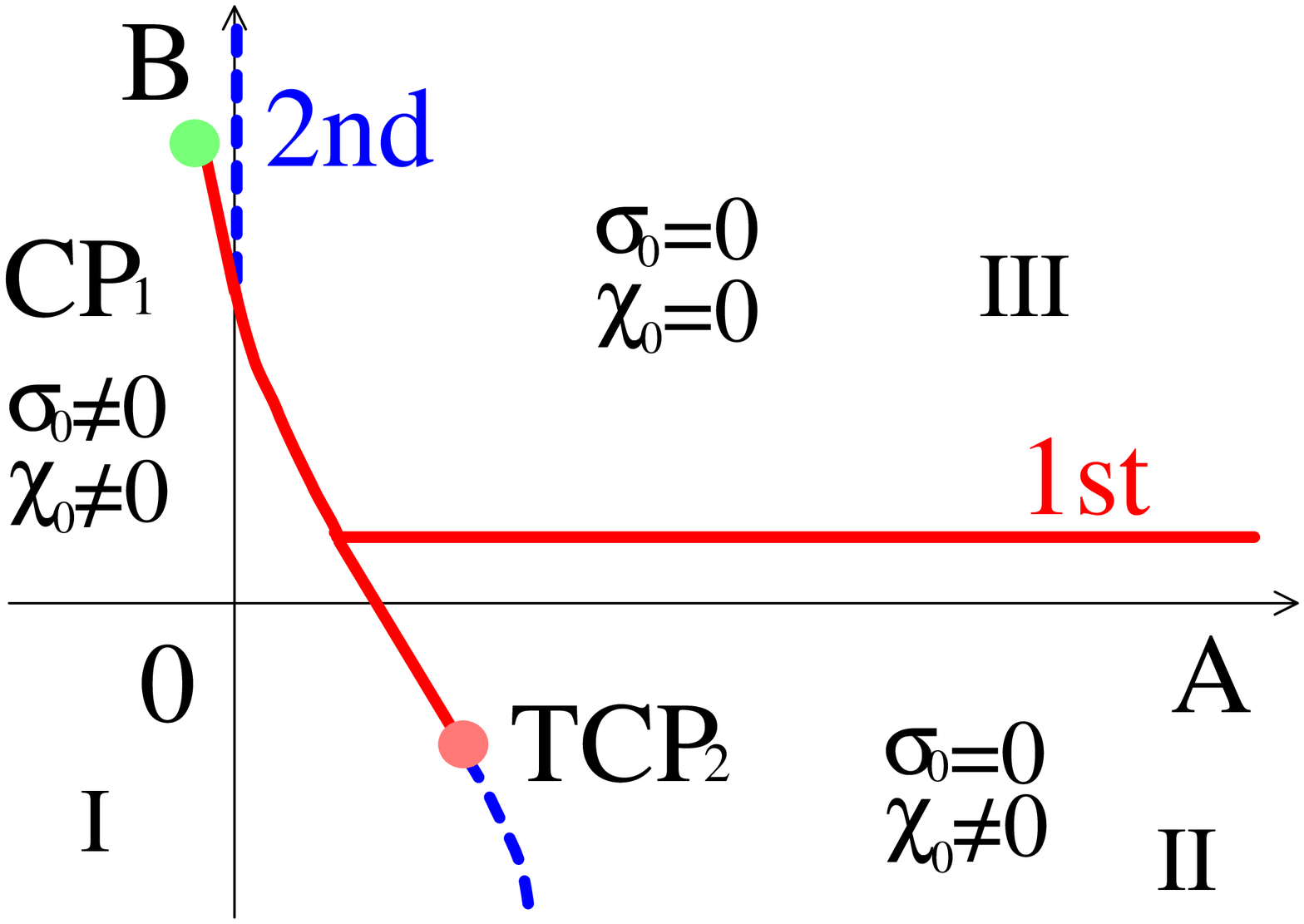}
\hspace*{1cm}
\includegraphics[width=7cm]{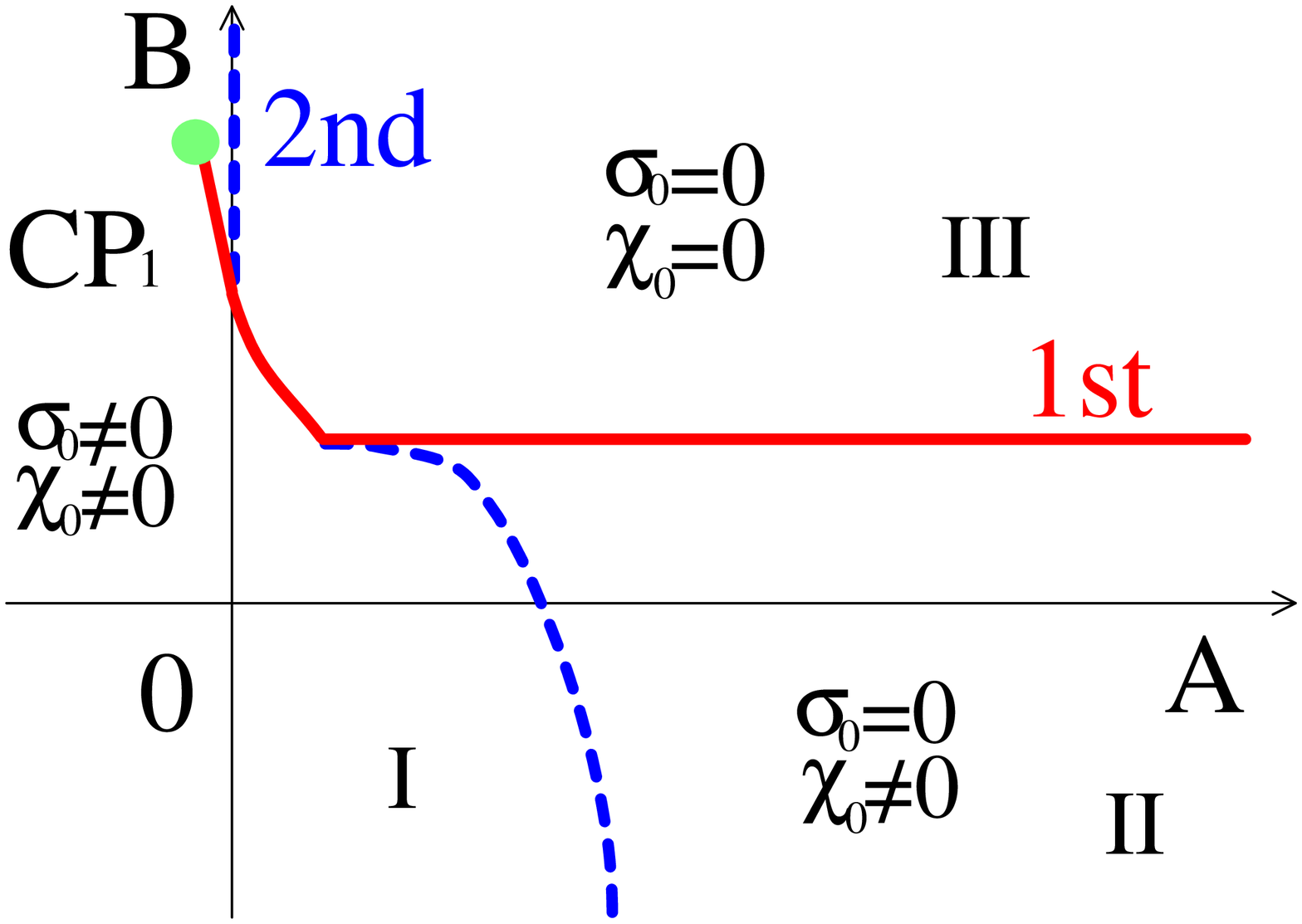}

\vspace*{0.5cm}

(iii) $0 < D < 1$
\hspace*{6cm}
(iv) $1 \leq D$
\\
\includegraphics[width=7cm]{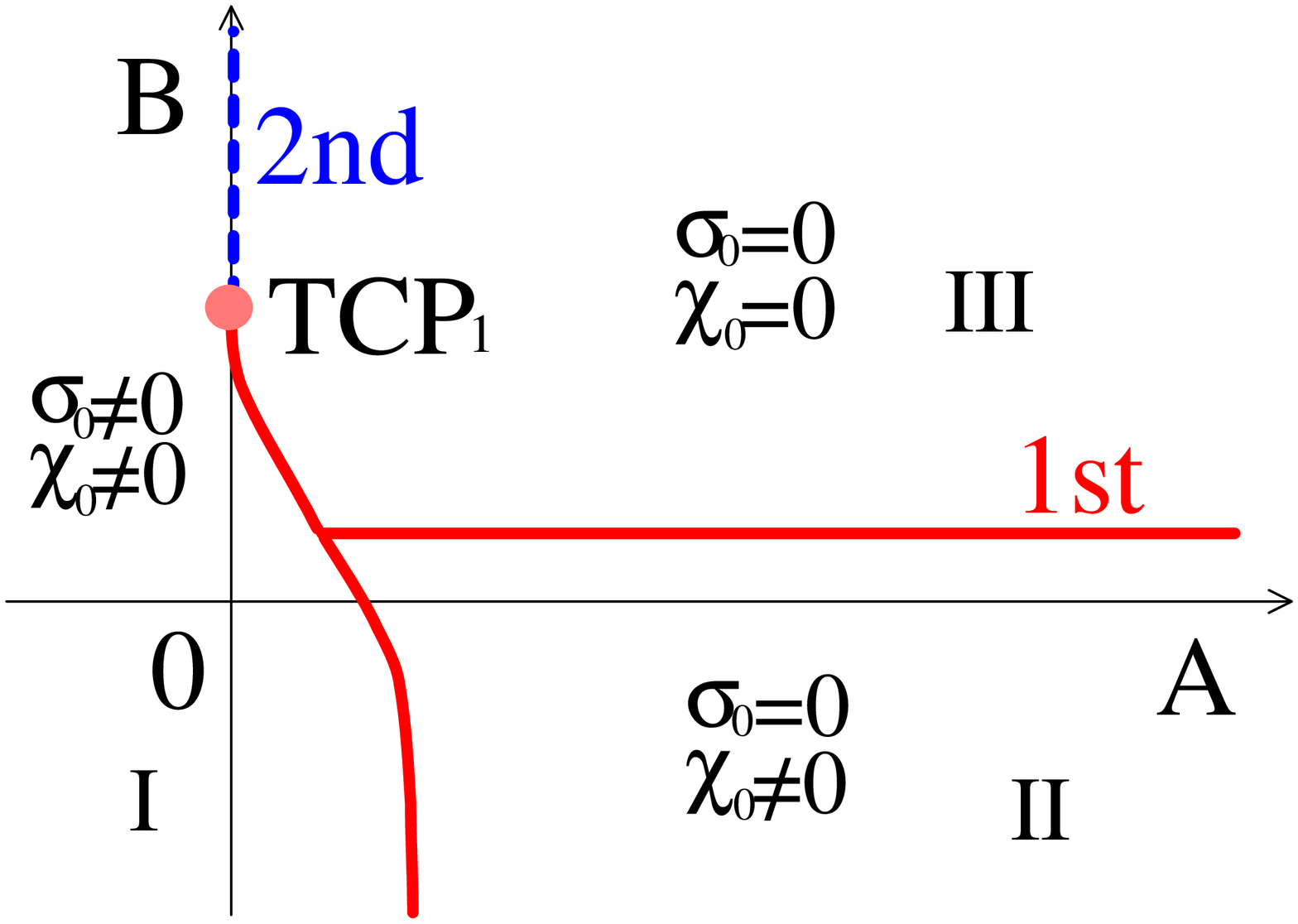}
\hspace*{1cm}
\includegraphics[width=7cm]{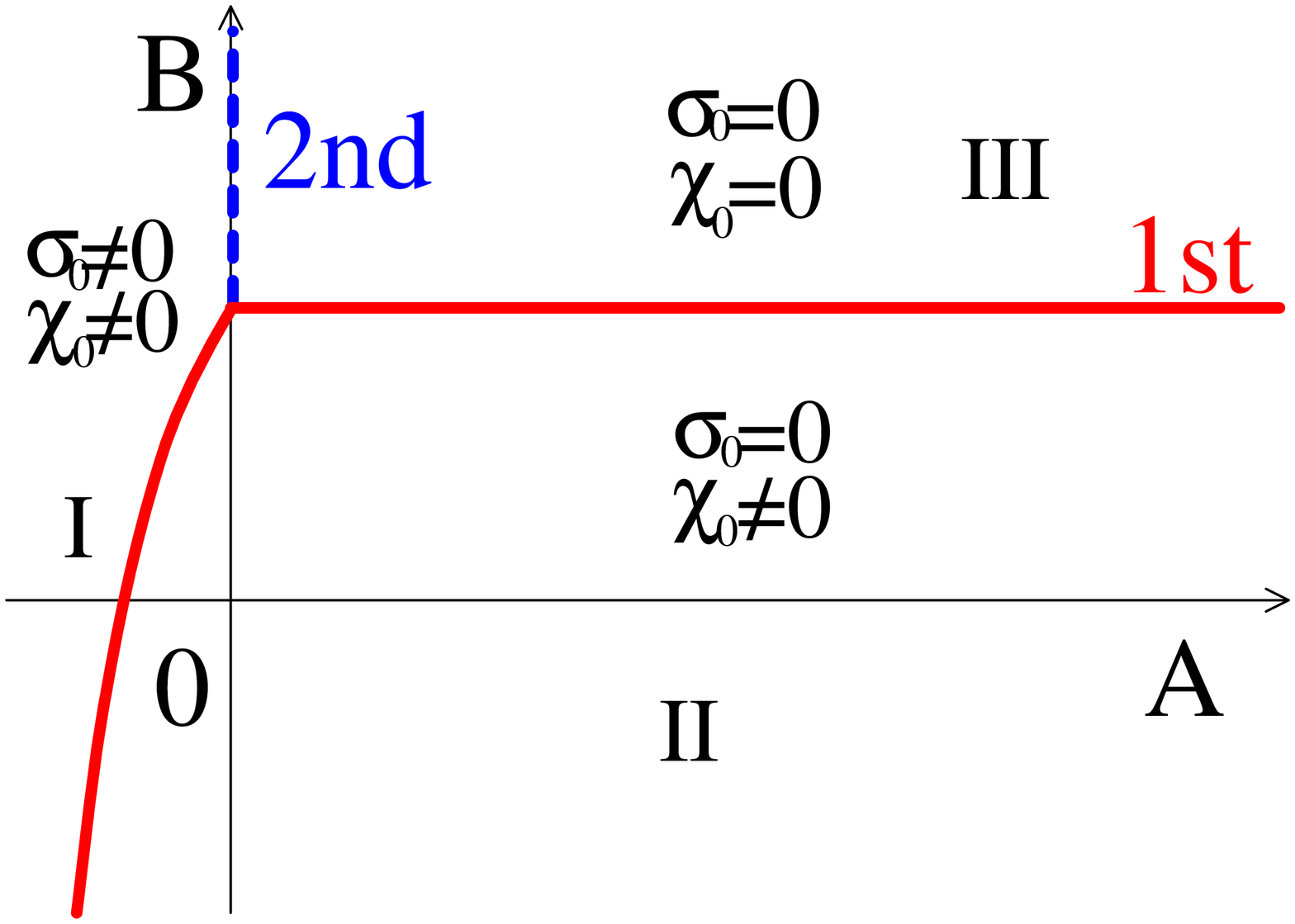}
\caption{
Phase diagram for different values of $D$
under $F=0$ and $h=0$.
The solid and dashed lines indicate first and second order 
phase boundaries, respectively.
}
\label{df}
\end{center}
\end{figure*}

A cubic term in $\chi$ modifies the previous phase structure
shown in Fig.~\ref{d0}.
The phase diagram from the potential,
\begin{equation}
V = A\sigma^2 + B\chi^2 + \sigma^4 + \chi^4 - \sigma^2\chi
{}+ D\chi^3\,,
\end{equation}
is classified by the following regions of $D$:
(i) $-1 < D < 0$\,,
(ii) $D \leq -1 $\,,
(iii) $0 < D < 1$ and
(iv) $1 \leq D$.
One observes a deformation of the boundary lines
depending on $D$ as in Fig.~\ref{df}.
The phase transition line separating phase II from 
phase III becomes of first order due to the presence of $D\chi^3$.
For any negative $D$, (i) and (ii), 
a critical point CP$_1$ appears as a remnant of TCP$_1$ for $D=0$.
TCP$_2$ remains on the phase diagram
for $-1 < D < 0$, (i), which
eventually coincides with the triple point
at $D=-1$, (ii).
For positive $D$, (iii) and (iv),
the transition line which separates phase I
from phase II turns to be of first order everywhere.
The triple point approaches 
the TCP$_1$ and coincides when a positive $D$ reaches unity.
The different order of phase transition
between phase I and phase II for $-1 < D < 0$ to
that for $0 < D <1$ can be understood as follows: 
For $D=0$ (see Fig.\ref{d0}) the vacuum expectation value (VEV) 
$\chi_0$ is positive in phase I near the phase boundary between 
phase I and II
due to the existence of the $-\sigma^2\chi$ term in the potential.
In phase II, on the other hand, when the positive $\chi_0$ 
provides a local minimum of the potential, $-\chi_0$ also does, 
and both coincide with the global minima.
These two vacua are physically equivalent, so that the phase 
transition from phase I to phase II can be of second order.
When we add $D\chi^3$ term with negative $D$ to the potential, 
the local minimum corresponding to the positive $\chi_0$ is only 
the global minimum in phase II.
This can be smoothly connected to the vacuum in phase I
where the VEV $\chi_0$ is positive.
On the other hand, when $D$ is positive, 
the negative $\chi_0$ gives the global minimum in phase II.
Thus, there is a mismatch of $\chi_0$ along the phase boundary
separating phase I from phase II, which indicates a first-order 
transition.

$D$ also affects the quark number susceptibility $\chi_q$.
As in the case of $D=0$, the $\chi_q$ exhibits a more relevant 
increase toward the $Z_2$ symmetry restoration than at
the chiral phase transition.
The critical exponents of $\chi_q$ is summarized in 
Table~\ref{qsusexp}.
\begin{table}
\begin{center}
\begin{tabular*}{7cm}{@{\extracolsep{\fill}}|c||ccc|}
\hline
{} & CP$_1$ & TCP$_1$ & TCP$_2$\\
\hline\hline
$D < 0$ & $2/3$ & --- & $1/2$\\
\hline
$D=0$ & --- & $2/3$ & $1/2$\\
\hline
$D > 0$ & --- & $1/2$ & --- \\
\hline
\end{tabular*}
\end{center}
\caption{
The critical exponents of the quark number susceptibility
for vanishing and non-vanishing $D$ at two tricritical points
and at the critical point (CP).
} 
\label{qsusexp}
\end{table}
One finds that the two regions, $D \leq 0$ and $0 < D$, corresponds
to different universality. The cubic term plays a similar role
to an explicit symmetry breaking term in the potential. This
may be an origin for the appearance of a critical point.

For  $-1 < D < 0$, TCP$_2$ for $h=0$ becomes a critical point, 
CP$_2$, for finite $h$.  
When the value of $h$ is increased, 
the CP$_2$ approaches the triple point
and coincides with it for a certain value of $h$, $h_0$.
The topology of the phase diagram for larger $h \geq h_0$ agrees
with that for $D \leq -1$. 
Similarly, the TCP$_1$ in the $0<D<1$ phase diagram becomes a critical
point CP$_1$ and disappears for a sufficiently large $h$. 
On the other hand, the CP$_1$ stays in the phase diagram Fig.~\ref{df} 
(i) and (ii) for any value of $h$.  
The scaling of $\chi_q$ there will be given by
\begin{equation}
\chi_q \sim t^{-2/3}\,.
\end{equation}

We note that 
adding finite $F$ to the potential does not generate
any essential differences from the above result with $F=0$.

\section{Hypothetical phase diagram and quark number susceptibility}
\label{sec:phase}

\begin{figure*}
\begin{center}
\includegraphics[width=7cm]{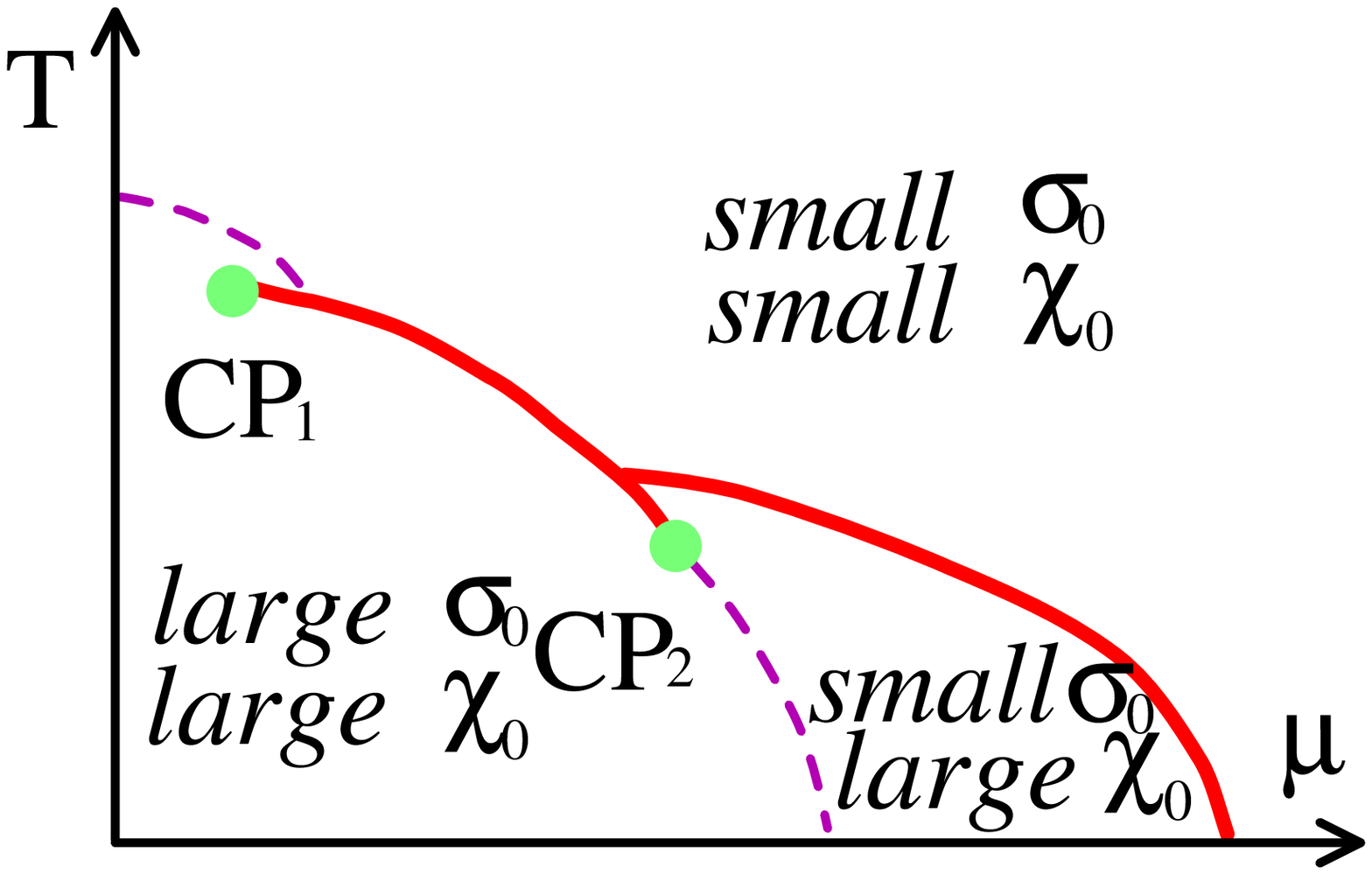}
\hspace*{1cm}
\includegraphics[width=7cm]{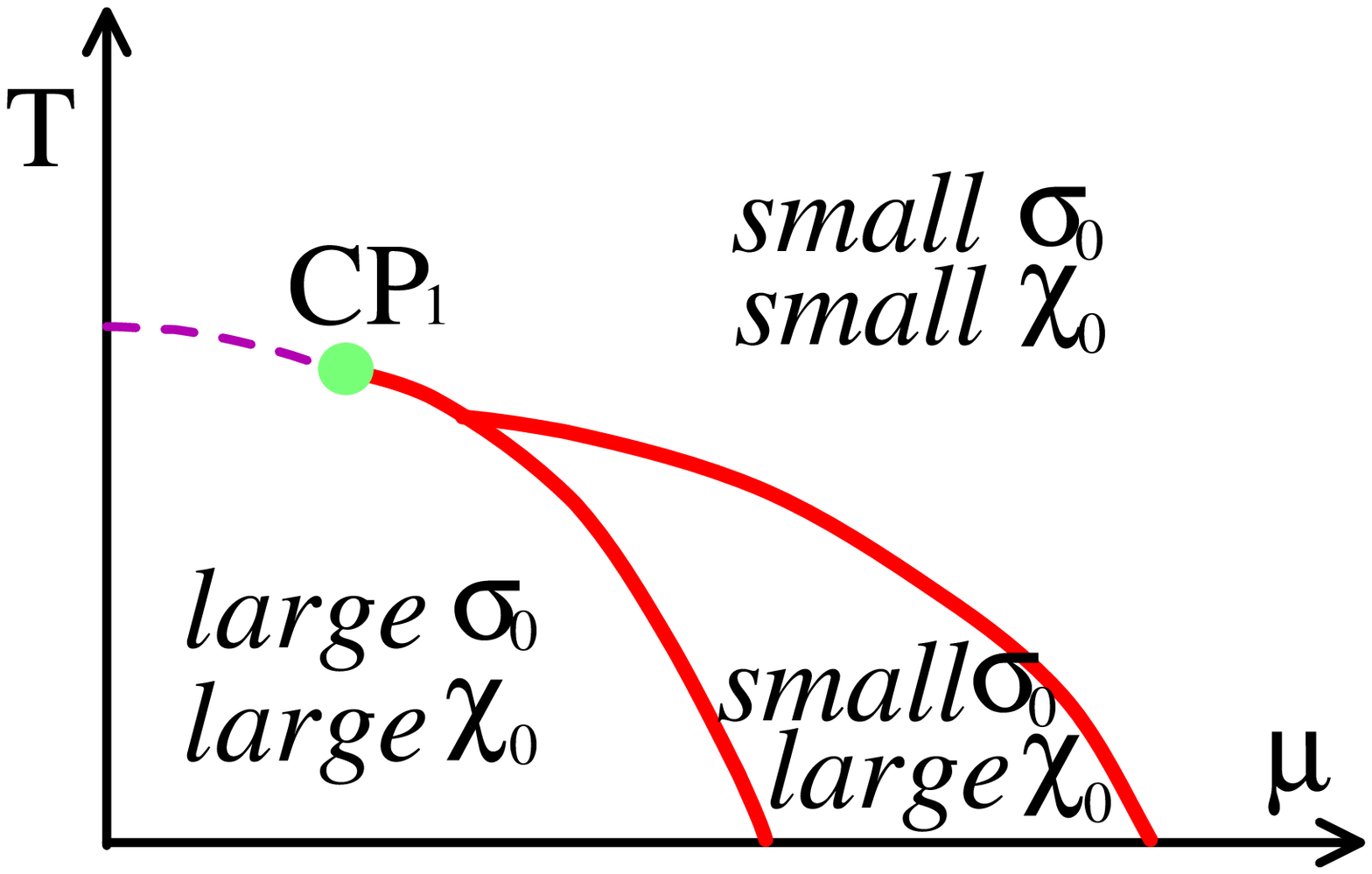}
\caption{
Schematic phase diagram mapped onto $(T,\mu)$ plane
with a negative $D$ (left) and with a positive $D$ (right).
The solid lines indicate first order phase boundaries, 
and dashed lines correspond to cross over.
}
\label{tmu}
\end{center}
\end{figure*}

From the above observations one would expect phase diagrams
mapped onto $(T,\mu)$ plane.
In the chiral limit a new phase where the center symmetry
is unbroken but chiral symmetry remains broken might appear
in dense matter since at $\mu=0$ this phase is strictly
forbidden by the no-go theorem.
With an explicit breaking of chiral symmetry
one would draw a phase diagram as in Fig.~\ref{tmu}.
The intermediate phase remains characterized
by a small condensation $|\sigma_0| \ll |\chi_0|$.
One would expect a new critical point associated with
the restoration of the center symmetry, CP$_2$, rather
than that of the chiral symmetry if dynamics prefers
a negative coefficient of the cubit term in $\chi$.
Multiple critical points in principle can be observed
as singularities of the quark number susceptibility.

It has been suggested that a similar critical 
point in lower temperature could appear in the QCD phase diagram
based on the two-flavored Nambu--Jona-Lasinio model with vector 
interaction~\cite{KKKN} and a Ginzburg-Landau potential with the 
effect of axial anomaly~\cite{yamamoto}.
There the interplay between the chiral (2-quark) condensate and 
BCS pairings plays an important role.
In our framework without diquarks, the critical point discussed 
in Fig.~\ref{tmu} (left) is 
driven by the interplay between the 2-quark and
4-quark condensates, and is 
associated with restoration of the center symmetry where 
anomalies have nothing to do with its appearance.
Nevertheless, the cross over in low temperatures may have
a close connection to the quark-hadron continuity~\cite{cont}
and it is an interesting issue to explore a possibility
of dynamical center symmetry breaking in microscopic calculations.
The present potential~(\ref{gl}) leads to a first-oder transition
of chiral symmetry even with an explicit breaking. This may be
replaced with a cross over when one considers higher order terms
in fields and other symmetry breaking terms as well as in-medium
correlations to baryonic excitations, which is beyond the scope
of this paper.

\begin{figure}
\begin{center}
\includegraphics[width=7cm]{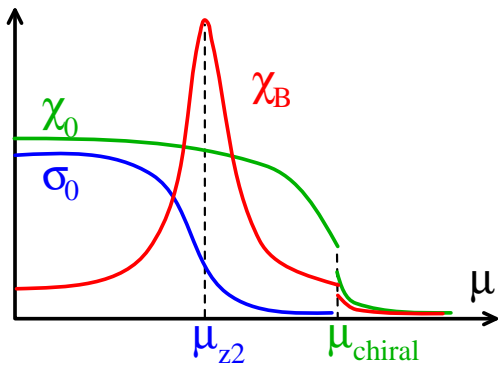}
\caption{
The behavior of the baryon number susceptibility
as a function of chemical potential assuming the phase
diagram of Fig.~\ref{tmu} (left).
The condensates and the susceptibility show
a jump also at $\mu_{z2}$ when the phase structure of
Fig.~\ref{tmu} (right) is preferred.
}
\label{qsus}
\end{center}
\end{figure}

Appearance of the above intermediate phase seems to have
a similarity to the notion of Quarkyonic 
Phase~\cite{quarkyonic,mrs:pnjl}, 
which is originally proposed as a phase of dense matter in large
$N_c$ limit. The transition from hadronic to quarkyonic world
can be characterized by a rapid change in the net baryon number
density. This feature is
driven by the restoration of center symmetry and is due to the 
fact that the Yukawa coupling 
of $\chi$ to baryons is not allowed by the $Z_2$ invariance.
Fig.~\ref{qsus} shows an expected behavior of
the quark (baryon) number susceptibility which exhibits
a maximum when across the $Z_2$ cross over.
This can be interpreted as the realization  
of the quarkyonic transition in $N_c=3$ world.
How far $\mu_{z2}$ from $\mu_{\rm chiral}$ is depends
crucially on its dynamical-model description.~\footnote{
 Thus, the present analysis does not exclude the possibility
 that both transitions take place simultaneously and in such
 case enhancement of $\chi_B$ is driven by chiral phase transition.
 The phase with $\chi_0\neq0$ and $\sigma_0=0$ does not seem to
 appear in the large $N_c$ limit~\cite{SDE2,shifman,fukushima}.
 It would be expected that including $1/N_c$ corrections induce 
 a phase with unbroken center symmetry.
}

\begin{figure}
\begin{center}
\includegraphics[width=7cm]{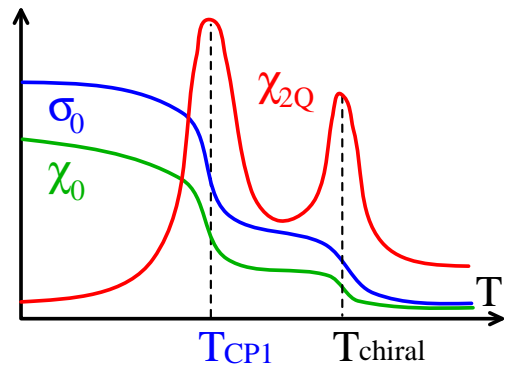}
\caption{
A schematic behavior of the susceptibility $\chi_{2Q}$
near the CP$_1$ as a function of temperature assuming 
the phase diagram of Fig.~\ref{tmu} (left).
}
\label{csus}
\end{center}
\end{figure}

It should be noticed that
the critical point in low density region,
indicated by CP$_1$ in Fig.~\ref{tmu} (left),
is different from a usually considered CP~\cite{CP}
in the sense that the CP$_1$ is not on the cross over 
line attached to the $T=0$ axis.
When we take a path from the broken phase 
(both $\sigma_0$ and $\chi_0$ are large) to the 
symmetric phase (both $\sigma_0$ and $\chi_0$ are small)
passing near the CP$_1$,  
the $\chi_{2Q}$ may exhibit two peaks;
one is located near CP$_1$ and another is on the
cross over line.
We show a schematic behavior of $\chi_{2Q}$ as a function of
temperature, together with $\sigma_0$ and $\chi_0$ in Fig.~\ref{csus}.
The appearance of two peaks in $\chi_{2Q}$ reflects the fact that 
$\sigma_0$ becomes small across the CP$_1$ and the cross over.
The first decrease in $\sigma_0$ near CP$_1$ is caused by a dropping 
$\chi_0$, while the second is by the chiral symmetry restoration.

\section{Hadron mass spectra and pion decay constant}
\label{sec:fpi}

In this section we derive meson mass spectra in a linear sigma model.
The Lagrangian with the potential~(\ref{potential}) is expressed
in terms of the mesonic fields as
\begin{eqnarray}
{\cal L}
&=& \frac{1}{2}\left( \partial_\mu\sigma\partial^\mu\sigma
{}+ \partial_\mu\vec{\phi}\cdot\partial^\mu\vec{\phi} \right)
\nonumber\\
{}&+& \frac{1}{2}\left( \partial_\mu\chi\partial^\mu\chi
{}+ \partial_\mu\vec{\psi}\cdot\partial^\mu\vec{\psi} \right)
{}- {\cal U}(\sigma,\phi,\chi,\psi)\,,
\end{eqnarray}
with
\begin{eqnarray}
&&
{\cal U}
= -\frac{m^2}{2}\left( \sigma^2 + \vec{\phi}\,^2 \right)
{}+ \frac{\lambda^2}{4}\left( \sigma^2 + \vec{\phi}\,^2 \right)^2
\nonumber\\
&&
{}- \frac{\bar{m}^2}{2}\left( \chi^2 + \vec{\psi}\,^2 \right)
{}+ \frac{\bar{\lambda}_1^2}{4}
\left[ \frac{1}{3}\chi^4 + \frac{2}{3}\chi^2 \vec{\psi}\,^2 
{}+ \frac{1}{2}\left(\vec{\psi}\,^2\right)^2 \right]
\nonumber\\
&&
{}+ \frac{\bar{\lambda}_2^2}{4}\left( \chi^2 + \vec{\psi}\,^2 \right)^2
{}- g\left[ 
  \frac{1}{2\sqrt{3}}\,\chi\left( 3\sigma^2-\vec{\phi}\,^2 \right) 
  {}+ \sqrt{2}\,\sigma \vec{\phi}\cdot\vec{\psi}
\right]
\nonumber\\
&&
{}+ 
\frac{g_3}{\sqrt{3}}\left( 
\frac{1}{3}\,\chi^3 + \frac{1}{2}\,\chi\vec{\psi}\,^2 
\right) \,,
\label{pot}
\end{eqnarray}
where $g_1 \equiv -g\,\,(g>0)$ and $g_2=0$ were taken.
In addition, we set $g_4=0$ since the $g_4$-term generates
only a shift in $m^2$ for $N_f=2$.
We also set the explicit breaking being zero.

The condensate of the mesonic fields 
in the phase where both the chiral symmetry and its center
$Z_2$ are broken 
are determined from the coupled gap equations given by
\begin{eqnarray}
\sigma_0^2 
&=& \frac{2}{\sqrt{3}\,g}
\left( 
 \frac{\bar{\lambda}^2}{3}\chi_0^2 - \bar{m}^2  
 + \frac{g_3}{\sqrt{3}} \chi_0
\right)
\chi_0\,,
\nonumber\\
\chi_0
&=& \frac{1}{\sqrt{3}\,g}\left( \lambda^2\sigma_0^2 - m^2 \right)\,,
\label{gap eq}
\end{eqnarray}
with $\bar{\lambda}^2 \equiv \bar{\lambda}_1^2 + 3\bar{\lambda}_2^2$.
Shifting the fields as
\begin{equation}
\sigma \to \sigma + \sigma_0\,,
\quad
\chi \to \chi + \chi_0\,,
\end{equation}
the potential reads
\begin{eqnarray}
{\cal U}
&=& 
\frac{1}{2}m_\sigma^2 \sigma^2
{}+ \frac{1}{2}m_\phi^2 \vec{\phi}\,^2
{}+ \frac{1}{2}m_\chi^2 \chi^2
{}+ \frac{1}{2}m_\psi^2 \vec{\psi}\,^2
\nonumber\\
&&
{}- \sqrt{3}\,g\sigma_0\, \sigma\chi
{}- \sqrt{2}\,g\sigma_0\, \vec{\phi}\cdot\vec{\psi}
{}+ \cdots\,,
\end{eqnarray}
where ellipses stand for the terms including the fields
more than three, and
\begin{eqnarray}
&&
m_\sigma^2 = 2\lambda^2\sigma_0^2\,,
\quad
m_\chi^2 = \frac{\sqrt{3}}{2}\frac{g}{\chi_0}\sigma_0^2
{}+ \frac{2}{3}\bar{\lambda}^2\chi_0^2
+ \frac{1}{\sqrt{3}}g_3\chi_0
\,,
\nonumber\\
&&
m_\phi^2 = \frac{4}{\sqrt{3}}\,g\chi_0\,,
\quad
m_\psi^2 = \frac{\sqrt{3}}{2}\frac{g}{\chi_0}\sigma_0^2\,,
\end{eqnarray}
The mass terms thus become
\begin{eqnarray}
{\cal U}^{(2)}
&=& 
\frac{1}{2} (\sigma\,,\chi)
\left(
\begin{matrix}
m_\sigma^2 & -\sqrt{3}\,g\sigma_0
\\
-\sqrt{3}\,g\sigma_0 & m_\chi^2
\end{matrix}
\right)
\left(
\begin{matrix}
\sigma
\\
\chi
\end{matrix}
\right)
\nonumber\\
{}&+&
\frac{1}{2} (\vec{\phi}\,,\vec{\psi})
\left(
\begin{matrix}
m_\phi^2 & -\sqrt{2}\,g\sigma_0
\\
-\sqrt{2}\,g\sigma_0 & m_\psi^2
\end{matrix}
\right)
\left(
\begin{matrix}
\vec{\phi}
\\
\vec{\psi}
\end{matrix}
\right)\,.
\end{eqnarray}
Obviously, the determinant of the above mass matrix for 
$\phi$ and $\psi$ is zero and thus massless pseudo-scalar
fields are a mixture of 2-quark and 4-quark states.

The mass eigenstates are introduced with a rotation matrix as
\begin{eqnarray}
&&
\left(
\begin{matrix}
S
\\
S^\prime
\end{matrix}
\right)
=
\left(
\begin{matrix}
\cos\theta & \sin\theta
\\
-\sin\theta & \cos\theta
\end{matrix}
\right)
\left(
\begin{matrix}
\sigma
\\
\chi
\end{matrix}
\right)\,,
\nonumber\\
&&
\left(
\begin{matrix}
\vec{P}
\\
\vec{P}^\prime
\end{matrix}
\right)
=
\left(
\begin{matrix}
\cos\bar{\theta} & \sin\bar{\theta}
\\
-\sin\bar{\theta} & \cos\bar{\theta}
\end{matrix}
\right)
\left(
\begin{matrix}
\vec{\phi}
\\
\vec{\psi}
\end{matrix}
\right)\,,
\label{rota}
\end{eqnarray}
with the angles
\begin{eqnarray}
\tan\left( 2\theta \right)
= \frac{2\sqrt{3}\,g\sigma_0}
{m_\chi^2 - m_\sigma^2}\,,
\quad
\tan\left( 2\bar{\theta} \right)
= \frac{4\sqrt{6}\,\sigma_0\chi_0}
{3\sigma_0^2 - 8\chi_0^2}\,.
\end{eqnarray}
The masses of scalar mesons are give by
\begin{eqnarray}
m_S^2 = m_\sigma^2\cos^2\theta + m_\chi^2\sin^2\theta
{}- \sqrt{3}\,g\sigma_0\sin(2\theta)\,,
\nonumber\\
m_{S^\prime}^2 = m_\chi^2\cos^2\theta + m_\sigma^2\sin^2\theta
{}+ \sqrt{3}\,g\sigma_0\sin(2\theta)\,,
\end{eqnarray}
and those of pseudo-scalar mesons by
\begin{eqnarray}
m_P = 0\,, \quad
m_{P^\prime}^2 = 
\frac{g(3\sigma_0^2 + 8\chi_0^2)}{2\sqrt{3} \chi_0}\,,
\label{masses pseudo}
\end{eqnarray}
with
\begin{equation}
\cos\bar{\theta} = \frac{ \sqrt{3}\sigma_0 }
  {\sqrt{ 3 \sigma_0^2 + 8 \chi_0^2 }}
\,,
\quad
\sin\bar{\theta} = \frac{ 2 \sqrt{2}\chi_0 }
  {\sqrt{ 3 \sigma_0^2 + 8 \chi_0^2 }}
\,.
\end{equation}
The pion decay constant is read from the Noether current, 
$J_A^\mu \sim \sigma_0\partial^\mu\phi 
{}+ 4/\sqrt{6}\,\chi_0\partial^\mu\psi$, as
\begin{equation}
F_\pi = \sqrt{\sigma_0^2 + \frac{8}{3}\chi_0^2}\,.
\end{equation}
Since we consider a system in the chiral limit, the
massive $P^\prime$ state is decoupled from the current
and $F_{\pi^\prime}=0$, as it should be.
It should be noted that, when $|\sigma_0| \gg |\chi_0|$,
the NG boson is dominantly the 2-quark state.
The 4-quark component becomes more relevant for
$\sqrt{3}|\sigma_0| < \sqrt{8}|\chi_0|$, i.e.
$\bar{\theta} > \pi/4$.

When the coupling $g_3$ is negative, which corresponds to
$D < 0$ in the Ginzburg-Landau potential given in 
section~\ref{sec:model},
the phase transition from phase I ($\sigma_0\neq0$ and 
$\chi_0\neq0$) to phase II ($\sigma_0 = 0$ and 
$\chi_0\neq0$) can be of second order.
In such a case,
the restoration of the center $Z_2$ symmetry is characterized
by vanishing $\sigma_0$. Approaching the restoration from
broken phase, one finds the lowest scalar meson mass
degenerate with the $P$ state,
while the pion decay constant remains finite
due to $\chi_0 \neq 0$;
\begin{eqnarray}
m_S \to m_P = 0\,,
\quad
F_\pi \to \sqrt{\frac{8}{3}}\,\chi_0\,,
\end{eqnarray}
with
\begin{eqnarray}
\chi_0 = \sqrt{\frac{3\bar m^2}{\bar\lambda^2}
+\left( \frac{\sqrt{3}\,g_3}{2\bar\lambda^2} \right)^2} 
- \frac{\sqrt{3}\,g_3}{2\bar\lambda^2} \,.
\end{eqnarray}

The vanishing $S$-state mass corresponds to a divergence of
the susceptibility $\chi_{\rm 2Q}$, which is responsible to
restoration of the center symmetry. 
The scalar $S$ and pseudo-scalar $P$ states thus become
the chiral partners on the phase boundary. 
In the $Z_2$ symmetric phase the 
meson masses are found from the potential~(\ref{pot}) as
\begin{eqnarray}
&&
m_\sigma^2 = - m^2 - \sqrt{3}\,g\chi_0
\,,
\quad
m_\phi^2 = - m^2 + \frac{g}{\sqrt{3}}\chi_0
\,,
\nonumber\\
&&
m_\chi^2 = \frac{2}{3} \bar{\lambda}^2 \chi_0^2
  + \frac{g_3}{\sqrt{3}} \chi_0
\,,
\quad
m_\psi^2 = 0\,.
\label{masses Z2}
\end{eqnarray}
There is no mixing in this phase, $\tan\theta=\tan\bar{\theta}=0$,
so that $\sigma$, $\phi$, $\chi$, $\psi$ are the mass 
eigenstates.~\footnote{
 When we approach the phase boundary from the
 $Z_2$ symmetric phase to the $Z_2$ broken phase,
 $m_\sigma^2$ in Eq.~(\ref{masses Z2}) approaches zero,
 since $-m^2 = \sqrt{3}g \chi_0$ is satisfied at
 the phase boundary.
 The pseudoscalar mass $m_\phi^2$ approaches
 $\frac{4}{\sqrt{3}} g \chi_0 $
 which coincides with the mass of $P'$ in the $Z_2$ broken phase
 (see Eq.~(\ref{masses pseudo})).
}
This implies that
the pure 4-quark state $\psi$ is the massless NG boson
in the $Z_2$ symmetric phase.
Due to the broken chiral symmetry, $\sigma$ and $\phi$ states
are not degenerate in mass.~\footnote{
 In reference~\cite{shifman} the degeneracy of 
 the massive scalar and pseudoscalar mesons
 made of 4-quarks carrying the same isospin
 for a general number of flavors was shown.
 In case of $N_f=2$ the U(1)$_A$ anomaly generates
 a mass difference between 
 the $\sigma$ state and 
 the pseudoscalar meson with $I=0$ ($\eta$).
 In the present analysis, we did not include 
 the $I=0$ 
 pseudoscalar and the $I=1$ scalar mesons
 from the beginning by 
 assuming that they are very heavy.
}
The vector and axial-vector
states neither degenerate in mass~\cite{shifman},
since both vector and axial-vector currents are invariant under
the $Z_2$ transformation but broken chiral symmetry does not
dictate the same masses.

When $| g_3 / g | \ll 1$, the chiral phase transition
from phase II ($\sigma_0=0$ and $\chi_0\neq0$)
to phase III ($\sigma_0=0$ and $\chi_0=0$) will be of
weak first-order.
In this case, $\chi_0$ then $F_\pi$ approach zero
near the phase transition point.
This is controlled by $\bar{m}$ approaching zero, 
which corresponds to $B$ approaching zero
in the Ginzburg-Landau potential 
discussed in section~{\ref{ssec:gl}}.
The iso-spin 2 state will become very light near the 
phase transition.
This may suggest that, when the $g_3\mbox{Det}\Sigma$ term is 
small and the chiral phase transition is of weak first-order,
a light exotic states with $I=2$ might exist in dense baryonic
matter.
When there exists the non-negligible $g_3\mbox{Det}\Sigma$ term,
on the other hand,
such state never becomes light since the chiral phase transition
is of strong first-order.

In two flavors, the system would prefer the parity doubling
for baryons 
in the $Z_2$ symmetric phase where the VEV $\chi_0$
does not generate the baryon masses~\cite{shifman}. 
In the parity doubling scenario~\cite{pdoubling},
all the baryons have their parity partners and then
each pair of parity partners has a degenerate mass.
On the other hand, in the naive scenario
the lightest baryon does not have a parity partner,
so that it becomes massless in the $Z_2$ symmetric phase.
We list hadron mass spectra expected in phase I and
phase II
in Table~\ref{mass}.
\begin{table*}
\begin{center}
\begin{tabular}{|l||l|}
\hline
\quad
phase I: $\sigma_0 \neq 0\,, \chi_0 \neq 0$ & 
\quad
phase II: $\sigma_0 = 0\,, \chi_0 \neq 0$  
\\
\hline\hline
\quad
$SU(2)_V$ & 
\quad
$SU(2)_V \times (Z_{2})_A$ 
\\
\hline
\quad
$m_S \neq 0\,, m_P = 0$ & 
\quad
$m_S \neq m_P \neq 0\,, m_{P^\prime} = 0$
\\
\quad
$m_V \neq m_A$ &
\quad
$m_V \neq m_A$ 
\\
\hline
\quad
$F_\pi = \sqrt{\sigma_0^2 + (8/3)\chi_0^2}$ &
\quad
$F_\pi = \sqrt{8/3}\,\chi_0$ 
\\
\hline
\quad
$m_{N^+} \neq 0$ &
\quad
(i) naive: 
$\left\{\begin{array}{l}
  m_{N^+} = 0 \mbox{(ground state)} \\
  m_{N^{\prime+}} = m_{N^{\prime-}} \neq 0 \\
  \quad \mbox{(excited states)}
\end{array}\right.$
\\
{} &
\quad
(ii) mirror: 
$\left\{\begin{array}{l}
  m_{N^+} = m_{N^-} \neq 0 \\
  \quad \mbox{(all states)}
\end{array}\right.$
\\
\hline
\end{tabular}
\end{center}
\caption{The mass spectra of mesons and baryons 
in different phases for $N_f=2$. Baryons transform with the naive
chirality assignment as $\psi_{R,L} \to g_{R,L} \psi_{R,L}$, while
with the mirror assignment as $\psi_{1R,L} \to g_{R,L} \psi_{1R,L}$
and $\psi_{2R,L} \to g_{L,R} \psi_{2R,L}$ with 
$g_{R,L} \in SU(2)_{R,L}$ where two nucleons $\psi_1$ and 
$\psi_2$ belong to the same chiral multiplets.
}
\label{mass}
\end{table*}

\section{Conclusions}

We have discussed a new phase where chiral symmetry is
spontaneously broken while its center symmetry is restored.
This might appear as an intermediate state
between chirally broken and restored phases in $(T,\mu)$ plane. 
The appearance of the intermediate phase with unbroken 
$Z_2$ also suggests a new critical point associated with
the center symmetry in low temperatures.
A tendency of the center symmetry restoration is carried by
the net baryon number density which shows a rapid increase
and this is reminiscent of the quarkyonic transition.
In this phase there exist the NG bosons and thus 
nothing prevents the anomaly matching. 
It has been shown that the anomaly matching conditions are
also valid in gauge theories at finite density~\cite{anomaly}.
This may suggest that the phase III is chirally restored
and deconfined.
The $U(1)_A$ symmetry remains broken and the heavy $\eta$ mass 
can be controlled with a certain anomaly coefficient.

There are subtleties in baryon masses since the 
existence of the center symmetry
does not immediately dictate the parity doubling
for a general number of flavors: 
Here we consider the case in massless three flavors.
The $g_3$-term in (\ref{potential}) now generates $\chi^8$
contribution, while the $g_4$-term does $\sigma^3$ one.
It follows that $D\chi^3$ is removed from (\ref{gl}) and another
cubic term $\sigma^3$ is added. 
Omitting the cubic term $\sigma^3$ results in
the same phase diagram as Fig.~\ref{d0} with two TCPs.
When the cubic term $\sigma^3$ is included, 
it is conceivable that phase II and phase III in 
Fig.\ref{df} are separated by a second-order phase boundary, 
which will become a first-order one when we take quantum fluctuations 
into account~\cite{PW}.
The topologies are expected to be quite similar to those
shown in Fig.~\ref{df}, so that
we expect a strong enhancement of the quark number
susceptibility at the $Z_3$ restoration point.
Differently from the case for $N_f=2$, the $\Sigma_{ab}$ field
is allowed to
couple to the octet baryon states as, e.g. 
$\bar{B}_a \Sigma_{ab} B_b$,
and the baryon number current couples to the $\chi$ state
which becomes massless at the chiral restoration point.
As a result, the quark number susceptibility might show another
peak at the chiral restoration.
Hadron masses in the $Z_3$ symmetric phase
are slightly different from those under $Z_2$ invariance:
In the mesonic sector the party partners are degenerate
and the degeneracy does not generally
occur in the baryonic sector~\cite{shifman}.
Following Ref.~\cite{shifman}, possible operators for the
baryons are expressed as
\begin{eqnarray}
B_L^1 = (q_L q_L q_R)_L \,,\,\,
B_L^2 = (q_L q_R q_R)_L \,,\,\,
B_L^3 = (q_L q_L q_L)_L \,,
\nonumber\\
B_R^1 = (q_R q_R q_L)_R \,,\,\,
B_R^2 = (q_R q_L q_L)_R \,,\,\,
B_R^3 = (q_R q_R q_R)_R \,,
\nonumber\\
\end{eqnarray}
where the color and flavor indices are omitted.
For the octet baryons,
the representations under the chiral 
SU(3)$_L \times$SU(3)$_R$ of these baryonic fields are assigned as
\begin{eqnarray}
B_L^1 \sim (\bar{3}\,,\,3)\,,\quad
B_L^2 \sim (3\,,\,\bar{3})\,,\quad
B_L^3 \sim (8\,,\,1)\,,
\nonumber\\
B_R^1 \sim (3\,,\,\bar{3})\,,\quad
B_R^2 \sim (\bar{3}\,,\,3)\,,\quad
B_R^3 \sim (1\,,\,8)\,.
\end{eqnarray}
When the $B^3$ is the lightest octet baryon,
which we call the naive assignment,
it is still massive in the $Z_3$ symmetric phase,
since the Yukawa coupling of the 4-quark state $\Sigma_{ab}$
is possible as, e.g. $\bar{B}_a \Sigma_{ab} B_b$.
When the lightest baryons are described by a combination of
$B^1$ and $B^2$, which we call the mirror assignment,
they are degenerate with each other in the
$Z_3$ symmetric phase.
We summarize these features in Table~\ref{mass3}.
\begin{table*}
\begin{center}
\begin{tabular}{|l||l|}
\hline
\quad
phase I: $\sigma_0 \neq 0\,, \chi_0 \neq 0$ & 
\quad
phase II: $\sigma_0 = 0\,, \chi_0 \neq 0$   
\\
\hline\hline
\quad
$SU(N_f)_V$ & 
\quad
$SU(N_f)_V \times (Z_{N_f})_A$ 
\\
\hline
\quad
$m_S \neq 0\,, m_P = 0$ &
\quad
$m_S = m_P \neq 0$, $m_{P^\prime} = 0$ 
\\
\quad
$m_V \neq m_A$ &
\quad
$m_V \neq m_A$ 
\\
\hline
\quad
$m_{N^+} \neq 0$ &
\quad
(i) naive: $m_{N^+} \neq 0$
\\
{} &
\quad
(ii) mirror: $m_{N^+} = m_{N^-} \neq 0$ 
\\
\hline
\end{tabular}
\end{center}
\caption{Same as in Table~\ref{mass} but for $N_f = 3$.
}
\label{mass3}
\end{table*}
The baryon masses crucially depend on a way of chirality
assignment. It would be an interesting issue to clarify this
within a more elaborated model.

The main assumption in this paper is a dynamical breaking
of chiral symmetry $SU(N_f)_L \times SU(N_f)_R$ down to
a non-standard $SU(N_f)_V \times {(Z_{N_f})}_A$ although
this seems to be theoretically self-consistent.
Calculations using the Swinger-Dyson equations or
Nambu--Jona-Lasinio type models with careful treatment
of the quartic operators may directly evaluate this reliability.
Besides, anomalously light NG bosons, $m_\pi^2 \sim {\cal O}(m_q^2)$,
could lead to an s-wave pion condensation as discussed 
in~\cite{pioncondensate}.
A calculation using the Skyrme model shows a similar intermediate
phase~\cite{skyrmion}. Although the above non-standard pattern
of symmetry breaking was not imposed in the Skyrme Lagrangian,
the result could suggest an emergent symmetry in dense medium.
This intermediate phase would be an intriguing candidate
of the quarkyonic phase if it could sustain in actual QCD 
at finite density and would lead to a new landscape of dense 
baryonic matter.

\subsection*{Acknowledgments}

We are grateful for stimulating discussions 
with W.~Broniowski, K.~Fukushima, L.~McLerran, K.~Redlich and M.~Rho.
The work of C.S. has been supported in part  
by the DFG cluster of excellence ``Origin and Structure of the 
Universe''.
M.H. and C.S. acknowledge partial support by the WCU project 
of the Korean Ministry of Educational Science and Technology
(R33-2008-000-10087-0) and the warm hospitality by the members
of Hanyang University where this work was initiated.
The work of M.H. and S.T. has been
supported
in part by the JSPS Grant-in-Aid for Scientific Research
(c) 20540262 and Global COE Program ``Quest for
Fundamental Principles in the Universe'' of Nagoya
University (G07).

\appendix

\setcounter{section}{0}
\renewcommand{\thesection}{\Alph{section}}
\setcounter{equation}{0}
\renewcommand{\theequation}{\Alph{section}.\arabic{equation}}

\begin{widetext}

\section{Phase boundaries from Ginzburg-Landau potential}
\label{app:phase}

The relevant expressions for the phase boundaries obtained 
from the potential~(\ref{gl}) are given below.
We will take $D=F=0$ and the chiral limit $h=0$.

\begin{itemize}
\item
Second-order phase transition when $B \geq 1/4$:
\begin{equation}
A = 0\,.
\end{equation}
The solutions for $\sigma$ and $\chi$ on this boundary are given by
\begin{equation}
(\sigma_0\,,\chi_0) = (0\,,0)\,.
\end{equation}

\item
First-order phase transition when $0 \leq B < 1/4$ and
$0 < A \leq 1/8$:
\begin{eqnarray}
A = \left( \frac{3}{8} - \sqrt{\left(\frac{3}{8}\right)^2
{}+ \frac{1}{2}\left( B - \frac{1}{4} \right)}\right)
\sqrt{-\frac{3}{8} - 2\left( B - \frac{1}{4} \right)
{}+ \sqrt{\left(\frac{3}{8}\right)^2
{}+ \frac{1}{2}\left( B - \frac{1}{4} \right)}}\,.
\end{eqnarray}
The solutions for $\sigma$ and $\chi$ on this boundary are given by
\begin{eqnarray}
(\sigma_0\,,\chi_0) 
&=& (0\,,0)\,,
\\
&& \left( 
\pm\left[ \frac{1}{2}\left( \frac{1}{8} 
{}+ \sqrt{\left(\frac{3}{8}\right)^2
{}+ \frac{1}{2}\left( B - \frac{1}{4} \right)}\right) 
\sqrt{-\frac{3}{8} - 2\left( B - \frac{1}{4} \right)
{}+ \sqrt{\left(\frac{3}{8}\right)^2
{}+ \frac{1}{2}\left( B - \frac{1}{4} \right)}}
\right]^{1/2}\,,
\right.
\nonumber\\
&&
\left.
\frac{1}{2}
\sqrt{-\frac{3}{8} - 2\left( B - \frac{1}{4} \right)
{}+ \sqrt{\left(\frac{3}{8}\right)^2
{}+ \frac{1}{2}\left( B - \frac{1}{4} \right)}}
\right)\,.
\end{eqnarray}

\item
First-order phase transition when $-1/8 < B < 0$ and
$1/8 < A < 1/4$:
\begin{equation}
A = \frac{1}{8} - B\,.
\end{equation}
The solutions for $\sigma$ and $\chi$ on this boundary are given by
\begin{equation}
(\sigma_0\,,\chi_0) = \left( 0\,,\pm\sqrt{\frac{-B}{2}} \right)\,,
\quad
\left( \pm\sqrt{\frac{1}{2}\left( B + \frac{1}{8}\right)}\,,
\frac{1}{4} \right)\,.
\end{equation}

\item
Second-order phase transition when $B \leq -1/8$ and
$A \geq 1/4$:
\begin{equation}
A = \sqrt{-\frac{B}{2}}\,.
\end{equation}
The solutions for $\sigma$ and $\chi$ on this boundary are given by
\begin{equation}
(\sigma_0\,,\chi_0) = \left( 0\,,\pm\sqrt{\frac{-B}{2}} \right)\,.
\end{equation}

\item
Second-order phase transition when $A > 1/8$:
\begin{equation}
B = 0\,.
\end{equation}
The solutions for $\sigma$ and $\chi$ on this boundary are given by
\begin{equation}
(\sigma_0\,,\chi_0) = (0\,,0)\,.
\end{equation}
\end{itemize}

\end{widetext}


\end{document}